\documentclass[5p]{elsarticle}
\pdfoutput=1
\usepackage{amssymb}

\journal{Composite Science and Technology}
\usepackage{graphicx}
\usepackage{natbib}
\usepackage{amsmath}
\usepackage{epsfig}
\usepackage{epstopdf}
\usepackage{geompsfi-pdf2}
\let\pdfximagebbox\undefined

\begin{document}

\begin{frontmatter}



\title{A semi-analytical model for the wrinkling of laminates during consolidation over a corner radius}
\author{T. J. Dodwell\fnref{label1}}
\author{R. Butler \corref{cor1}\fnref{label1}}
\author{G. W. Hunt \fnref{label1}}

\cortext[cor1]{Corresponding Author: \texttt{r.butler@bath.ac.uk}}
\address{Department of Mechanical Engineering, University of Bath, Bath, BA2 7AY.\fnref{label1}}

\begin{abstract}
If carbon fibre layers are prevented from slipping over one another as they consolidate onto a non-trivial geometry, they can be particularly susceptible to {\em wrinkling}/{\em buckling} instabilities. A one dimensional model for wrinkling during consolidation over an external radius is presented. Critical conditions for the appearance of wrinkles provide possible manufacturing and design strategies to minimise wrinkling. Numerical results for the unstable post buckling paths show localized buckling behaviour which demonstrates qualitative agreement with micrographs of wrinkles.
\end{abstract}


\end{frontmatter}


\section{Introduction}\label{sec:Intoduction}

Whilst the basic advantages of composite laminates are well proven, they are often compromised by
high costs, long development time, and poor quality due to multiple defects, particularly in massive
complex parts such as those found in aerospace applications. The modelling, simulation and optimisation of manufacturing processes therefore has widespread applications to the industry, with the twin objectives of improving product quality and decreasing production time.

\subsection{Wrinkling of carbon fibre composites during consolidation}
\begin{figure*}
\centering
\includegraphics[width= 3in]{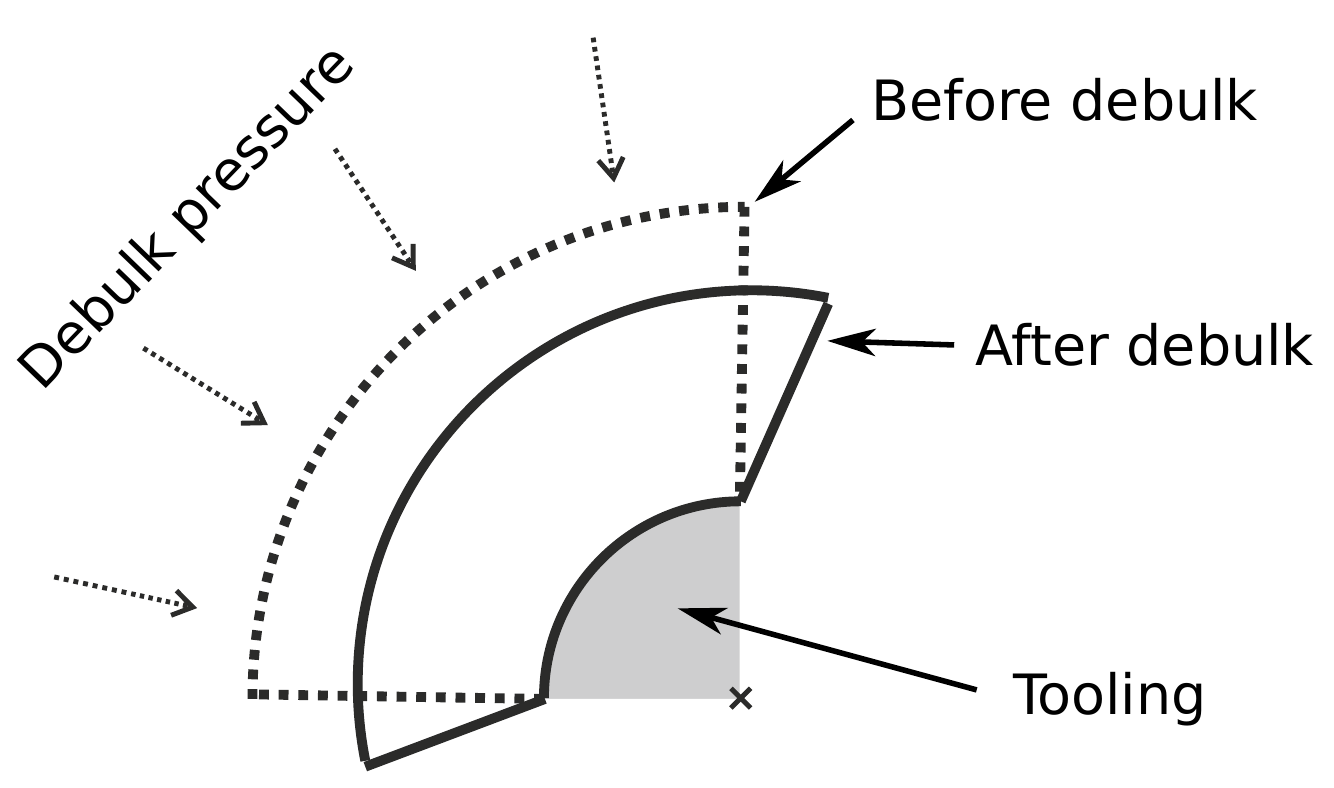}\includegraphics[width= 2.6in]{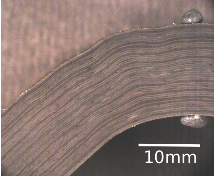}
\caption{(Left) A representation of the {\em bookend} effect, created when a laminate is consolidate over a corner radius and layers are free to slip over one another. (Right) A micrograph of small amplitude wrinkles in a typical corner radius. The wavelength, $\xi$ , of the outside ply is approximately $13$ mm with an amplitude of $0.7$ mm.}
\label{fig:introduction}
\end{figure*}

Typically, carbon fibre composite parts are made by layering a series of thin carbon fibre layers, pre-impregnated with resin, onto a tool surface. During this lay-up process the stack of plies is consolidated at moderate temperatures and pressures to remove air trapped between layers. This debulking process aims to ensure correct seating onto the tool surface, and to promote adhesion between plies. However, as a laminate consolidates over even a simple geometry the plies are forced to accommodate the imposed geometry of the tool surface. For example, consider consolidation over an external radius Fig.~\ref{fig:introduction} (left). As the outermost ply consolidates it is forced into a tighter geometry; if the layers cannot slip, they are put into axial compression. For plies in which the fibres align with this stress, their stiffness is particularly high ($\sim 230$ GPa~\cite{HexplyM21}). If layers can slip over one another the additional length can be accommodated by producing so called `book-ends', but if the resistance 
to slip is too high, layers may form wrinkles. Figure~\ref{fig:introduction} (right) shows small amplitude wrinkles or folds, which have developed in the corner radius of a large scale component. Understanding how wrinkles form during these manufacturing processes is important because, depending on their severity, they may compromise the structural integrity of the final part, leading in some cases to expensive wholesale rejection. The formation of wrinkles not only disrupts the even distribution of fibre and resin, but most significantly can increase through-thickness stresses, causing delaminations. This can trigger failure at significantly reduced loads~\cite{Adams1994}.

\subsection{Complexities of modelling layered systems}

\smallskip\noindent
A variety of models have been proposed for the consolidation of composite laminates. Typically, these take the form of flow-compaction continuum models, which couple a nonlinear elastic response of the fibres with a Darcy-type flow model for the redistribution of resin throughout the laminate~\cite{Gutowski1989,Hubert1998,Li2002}. However, with finely-layered structures and uncured laminates, slip in the interfaces between layers can introduce highly nonlinear, anisotropic behaviour. Rapidly-varying shear stresses though layer thickness, for example, can result from plies slipping and bending as individual layers rather than a combined composite. Current process models do not account for the anisotropy introduced by the layering, and as a result such models cannot encapsulate layer-level phenomena such as wrinkling.

\smallskip
To include the mechanics of individual layers, explicit finite element calculations can be performed using special interface elements~\cite{Wriggers2006}. Any number of interfaces could be modelled this way, yet such approaches are naturally restricted since mesh sizes must be sufficiently small compared with layer thickness. Some modelling based approaches have sought to include interlayer mechanics by deriving homogenised anisotropic continuum models~\cite{Biot1965}. Such models are effective if shear properties of the interface and the layer are similar, as for example in a cured laminate. However, for larger disparities, where the layers have the potential to undergo finite slip and separation, such models break down. For these cases, models must not only consider the anisotropic nature of shear at the interfaces, but also the individual contributions of layers as they bend. An {\em alternative approach}, taken here, is to incorporate individual contributions of both layer bending and work done in inter-
layer slip into a variational formulation~\cite{Dodwellthesis2012}. Here the interlayer geometry can be described by front propagation techniques such as the {\em level set method}~\cite{Boon2007}, or by assuming simplified interlayer relationships~\cite{Hunt2012b}. 

\subsection{Overview of the paper and comments on modelling}
This investigation is the first part of a more general study of the deformation of multi-layered uncured laminates. It has two distinct aims. First, the model (section~\ref{sec:Model}), has been created to understand the underlying elastic mechanisms which cause wrinkling during consolidation. Whilst the true response may result from a complex mix of nonlinear geometry and viscoelastic, temperature-dependent rheology, our objective here is to complement other high fidelity FE studies~\cite{Hubert1999,Li2002} with a simplified model, in which the parameters influencing the localised bucking can be closely defined and carefully monitored. By understanding the key parameters which control the formation of wrinkles (section~\ref{sec:criticalload}), the model aims to influence both design and manufacturing decisions, with the objective of reducing the possibility of wrinkles (section~\ref{sec:manufacturing}).

\smallskip
A second aim is to highlight new challenges which arise in modelling the complex behaviour of layered systems. If macroscopic models are to encapsulate localised wrinkling of individual layers within a multilayer material, constitutive descriptions must include localized contributions of individual layers in bending. The modelling of uncured laminates requires a shift of emphasis from the well-documented modelling of cured laminates. The relative weakness of the interfaces in shear/slip gives individual layers extra freedoms to deform independently. As a consequence thinly layered structures are extremely susceptible to {\em localised} buckling/wrinkling.

\section{The wrinkling model}\label{sec:Model}
The model comprises a stack of $N$ plies of uniform initial thickness $t$ and unit width, that have been laid over a tool surface characterised by the circular arc  $x_t = R_t\theta$, for $\theta$ in the range $[-\pi/4,\pi/4]$, and straight limbs of length $L$, see Figure~\ref{fig:setup}(i). The $i^{th}$ layer, numbering from the outside inwards, is described by a radius of curvature $R_i$ with arc-length parameter $x_i$ and total length $\ell_i = \frac{1}{2}\pi R_i$. All plies are assumed to be identical and inextensible in the fibre direction.

The modelling process is broken into two stages: first the {\em elastic buckling problem} and secondly the {\em viscous shearing of limbs}. We consider each step in turn, and then combine them to derive a critical condition for the formation of a wrinkle.

\subsection{The buckling problem}
\begin{figure*}
\centering
\includegraphics[width = 5in]{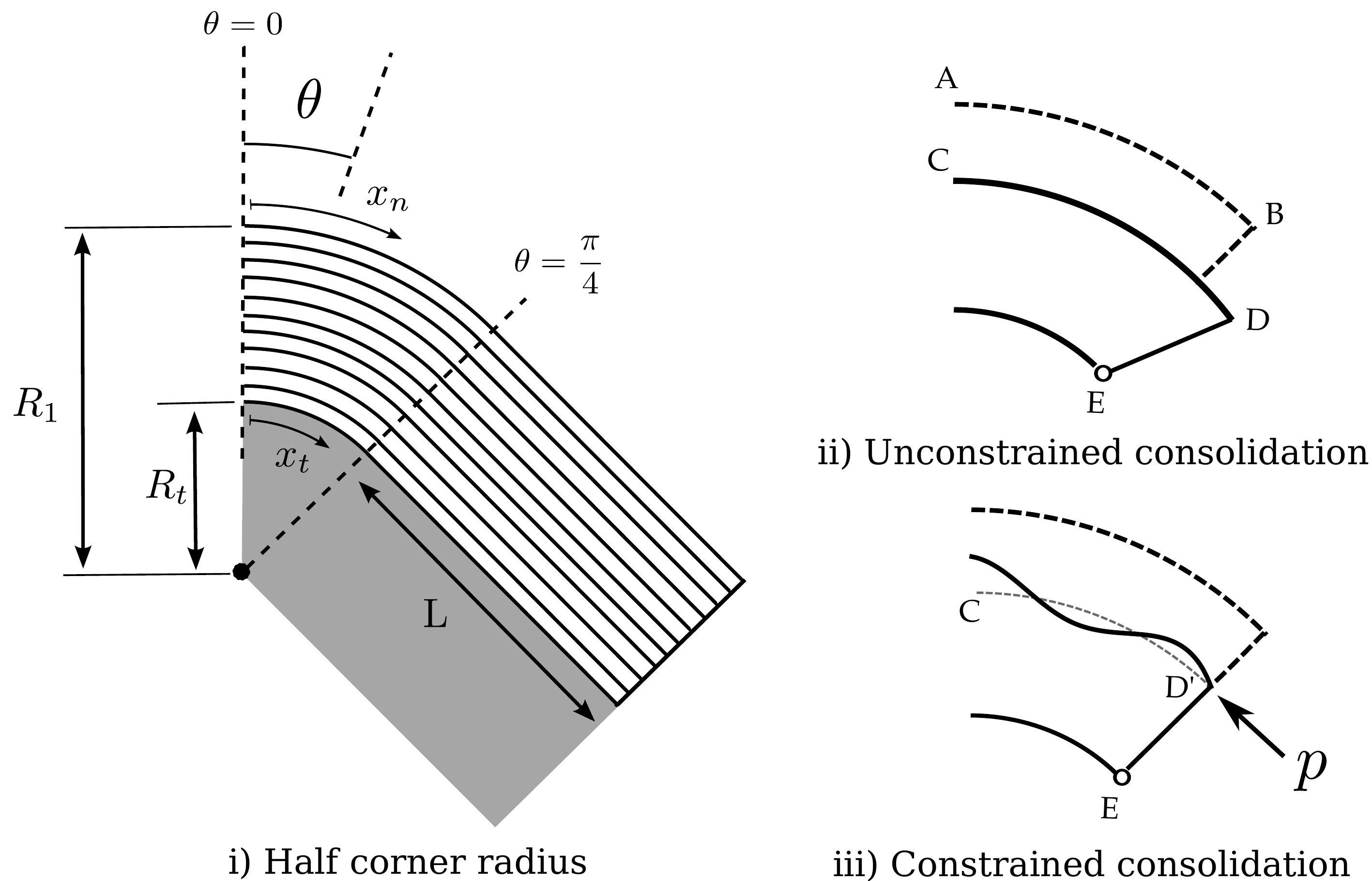}
\caption{(i) Setup of half the corner radius, identifying a number of key geometric parameters. (ii) and (iii) show two possible scenarios as a laminate consolidates over a corner radius. If, as in (ii), layers are free to slip over one another, {\em book-ends} form ($CDE$). If this slip is prevented, as in (iii), the inextensible laminate must wrinkle ($CD'E$).}
\label{fig:setup}
\end{figure*}
A uniform debulk pressure $q$ is first applied to the outside layer, causing the laminate to consolidate by a percentage $\eta$. This then imposes the loading for the wrinkling or buckling problem. The initial consolidation of the laminate is assumed to exhibit a general nonlinear stiffening behaviour~\cite{Gutowski1989} given by the power law
\begin{equation}
q(\eta) = C\eta^2,
\label{eqn:consolidation}
\end{equation}
for a constant $C>0$. The radius of curvature of the $i^{th}$ layer therefore reduces during consolidation to
\begin{equation}
R_i(\eta) = R_t + \frac{1}{2}t(1 - \eta)(2(N-i) + 1).
\end{equation}
The change of variable, ${\rm d}x_i = ({R_i}/{R_t})\,{\rm d}x_t$, proves useful throughout the analysis since the tool surface $x_t$ remains unchanged during consolidation.
\smallskip
Consider the two scenarios depicted in Figure~\ref{fig:setup}(ii) and (iii). If the layers are allowed to freely slip over each other, the original length $AB$ must remain equal to the length $CD$, and as a result book-ends will form, as in Figure~\ref{fig:setup}(ii). Next consider the alternative case of Figure~\ref{fig:setup} (iii) where layers are constrained from slipping over one another. Rigid load $p_i$ acting on layer $i$ prevents it from slipping a distance 
\begin{equation}
\lambda_i(\eta) = \frac 14 t\pi\eta \left(2(N-i)-1 \right),
\label{eqn:lambda}
\end{equation}
and causes it to buckle. We note that $\lambda_i$ varies linearly through thickness. The wrinkle deformation of each layer is characterised by the function $w_i(x_i)$, which measures the displacement of the $i$-th layer away from radius $R_i$ in the normal direction. 
\smallskip
Motivated by the micrograph of Figure~\ref{fig:introduction}, we assume the wrinkle deformation of the layer decays to zero at the tool face, with the amplitude varying as the square root of the distance from the tool surface. Consequently the complete deformation of the laminate can be written in terms of the displacement of the outermost layer, characterised by a single function $w = w_1$, where
\begin{equation}
w_i =  \sqrt{\frac{2(N-i) + 1}{2N -1}}w.
\label{eqn:lineardecay}
\end{equation}
This means that the setup of the model can be simplified to a single layer of effective elastic stiffness $\hat B$ (section~\ref{sec:bending}) laid around the corner radius $R(\eta)=R_1$, as seen in Fig.~\ref{fig:setup} (i). Wrinkle deformations $w$ away from $R$ are resisted by a nonlinear Winkler foundation which acts strictly locally and normal to the layer. This encapsulates the work done in wrinkling, as the material squashes internally, and is connected to the consolidation law~\eqref{eqn:consolidation}, (section~\ref{sec:foundation}). A single rigid load $p$ mimics the combined effect of all loads $p_i$, preventing them from slipping. 

\subsection{Shearing of the limbs}

\begin{figure*}
\centering
\includegraphics[height=2in]{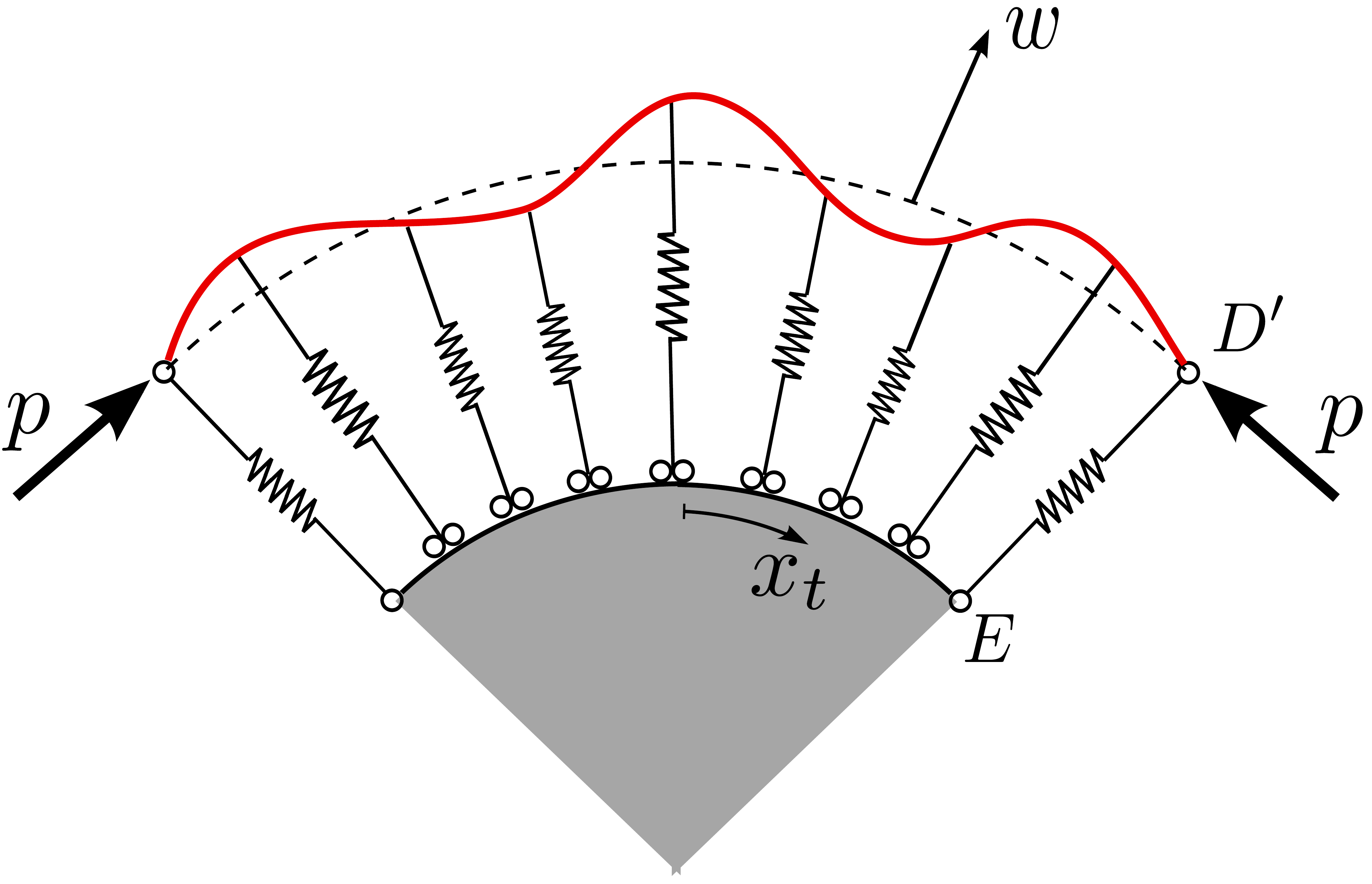}\includegraphics[height=2in]{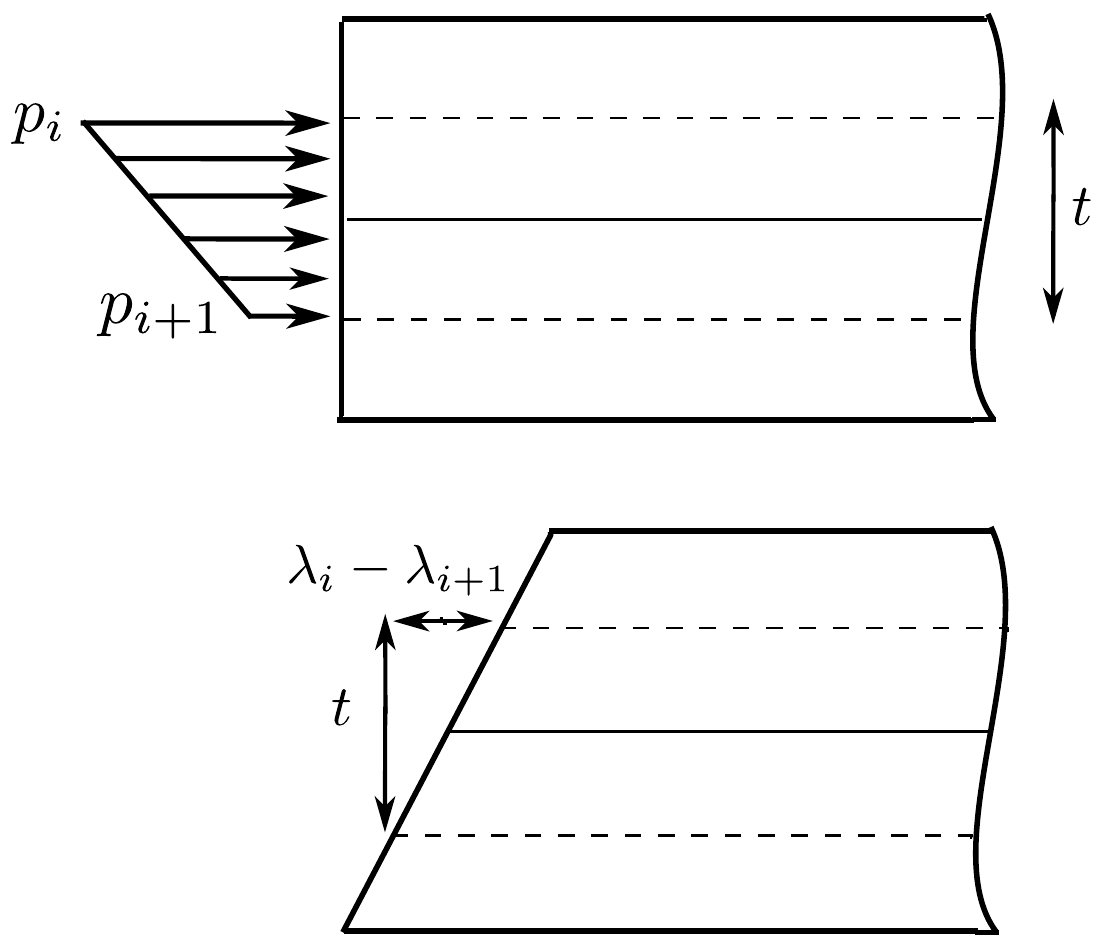}
\caption{Setup of the buckling model, simplified to a single layer of effective stiffness $\hat B$, laid around the corner radius $R(\eta)$, and on a nonlinear Winkler foundation. A rigid load $p$ prevents the laminate from producing {\em book-ends} and causes the laminate to wrinkle.}
\label{fig:model}
\end{figure*}

Accommodation of the bookend effect in  the limbs is taken to be dominated by the shearing properties of resin. The resin is modelled as a linear incompressible viscous material characterised by a viscosity $\mu$. In accordance with the two-stage assumption mentioned above, it is first assumed that at time  $T>0$ after initial consolidation,  the state is fully constrained as seen in Figure~\ref{fig:setup}(iii). The shear strain $\gamma$ at this position is given by $\gamma = \tau T/\mu$, where the shear stress $\tau = (p_i - p_{i+1})/t$ is the differential load acting on two adjacent layers divided by the layer thickness, $t$. By considering moments about $E$, (see Figure~\ref{fig:model}), $p$ can be related to the loads acting on each layer $p_i$, as follows:
\begin{equation}
p\left(R_1 - R_t\right)= \sum_{i=1}^{N}p_i\left(R_i-R_t\right),
\end{equation}
where loads $p_i$ vary linearly through the thickness such that
\begin{equation}
p_i = p_1\frac{2(N-i) + 1}{2N -1}.
\label{eqn:diffload}
\end{equation}  
The shear strain, $\gamma$, between two adjacent plies of length $L$ is $L(\lambda_i - \lambda_{i+1})/t$, which follows from~\eqref{eqn:lambda}. Therefore equilibrium is given by
\begin{equation}
\frac{\lambda_i - \lambda_{i+1}}{t}L = \frac{p_i - p_{i+1}}{\mu t}T.
\label{eqn:shearequil}
\end{equation}
\subsection{Critical wrinkling condition}
The two parts to the model are now combined to determine whether the buckling load can be accommodated by shear in the limbs. First, the buckling load $p$ is calculated; this then implies a differential load $p_i - p_{i+1}$, acts on two adjacent layers. Rearranging~\eqref{eqn:shearequil} for $L$ gives a critical limb length for a given debulk time $T$,
\begin{equation}
L_{crit} = \frac{1}{\mu}\frac{p_i - p_{i+1}}{\lambda_i - \lambda_{i+1}}T. 
\label{eqn:yield}
\end{equation}
For limbs lengths $L > L_{crit}$, a wrinkle will form.

\subsection{Energy-based formulation of the buckling problem}
\smallskip\noindent
We now construct the total potential energy from which we can derive a radial equilibrium equation for the model.

\subsubsection{Bending energy}\label{sec:bending}
\begin{figure}
\centering
\includegraphics[width=3.5in]{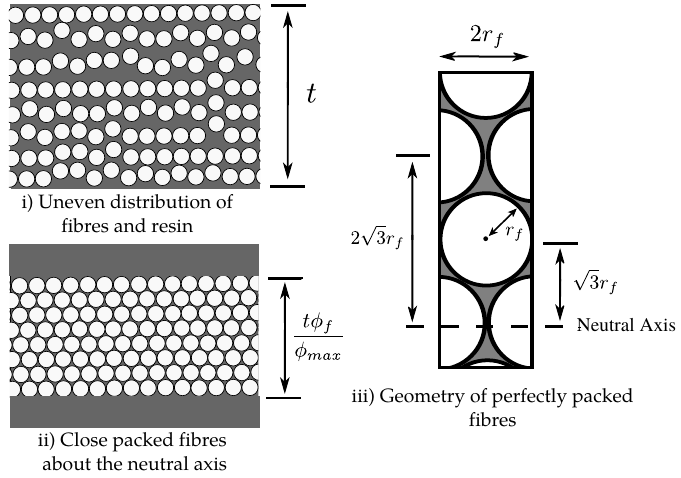}
\caption{(i) Ply of thickness $t$, with fibre volume fraction $\phi_f$, yet with complex local distributions of fibres and resin. (ii) Hexagonally-packed fibres with local fibre volume fraction $\phi_{max}$. (iii) Geometry of hexagonally-packed fibres.}
\label{fig:hexpacked}
\end{figure}

The relative weakness of the resin might be expected to lead to a complex redistribution of the fibres as each ply bends (see Figure~\ref{fig:hexpacked}(i)). To model the associated micro-mechanics is beyond the scope of this contribution. Here a conservative approximation is made for the elastic bending stiffness, by assuming that the fibres remain hexagonally close-packed as in Figure~\ref{fig:hexpacked}(ii). By considering an equilateral triangle of area $\sqrt{3}r_f^2$, consisting of hexagonally closed-packed fibres, the local fibre volume fraction of perfectly packed fibres is calculated to be $\phi_{max} = \sqrt{3}\pi/6$ (Figure~\ref{fig:hexpacked}(iii)). Arranged this way, the fibres occupy a reduce thickness $t\phi_f/\phi_{max}$ (Figure~\ref{fig:hexpacked}(ii)). If the contribution of resin to bending is neglected, the strain energy stored in bending for the i$^{th}$ layer is
\begin{equation}
U_{B_i} \simeq \frac{1}{2}E_f\underbrace{\int^{\frac{1}{2}t}_{-\frac{1}{2}t}z^2\;dz_f}_{:= I}\int^{\frac{1}{2}\ell_i}_{-\frac{1}{2}\ell_i}\kappa_i^2\;dx_i.
\end{equation}
where $\kappa_i$ defines the curvature of the $i^{th}$ layer. The second moment of area $I$ of the layer can be calculated by summing the contribution of each fibre. Assuming that one fibre lies directly on the mid-plane, the number of fibres through thickness is $2M + 1 = t\phi_f/(\phi_{max}r_f\sqrt{3})$. By noting that the centre of the i$^{th}$ fibre is offset a distance $z = ir_f\sqrt{3}$ from the neutral axis, it follows
\begin{align}
I &= \frac{1}{2r_f}\left[\frac{1}{4}\pi r_f^4 + 2\sum_{i=1}^M \frac{1}{4}\pi r_f^4 + 3\pi r_f^4 i\right] \\ \nonumber
&= \frac{1}{8}\pi r_f^3(2M+1)\left(1 + 4M(M+1)\right).
\label{eqn:SecondMoment}
\end{align}
Since the wavelengths of the wrinkle deformations are relatively small, changes in curvature of the mid-surface of $i^{th}$ layer can be approximated by $\kappa_i \simeq w^{\prime\prime}_i,$~\cite{Thompson1973}. The strain energy stored in bending is therefore the sum of the bending energy of each layer,
\begin{equation}
U_B = \frac{1}{2}E_fI \sum_{i=1}^{N} \int^{\frac{1}{2} \ell_i}_{-\frac{1}{2} \ell_i} w^{\prime\prime\; 2}_i\;dx_i
\end{equation}
By applying (\ref{eqn:lineardecay}) and noting that ${\rm d}x_i = ({R_i}/{R_t})\,{\rm d}x_t$ it follows that
\begin{equation}
U_B = \frac 12 \hat B \int^{\frac{1}{2} \ell_t}_{-\frac{1}{2} \ell_t} w^{\prime\prime\; 2}\;dx_t \; \mbox{where} \; \hat B= \frac{E_fI}{R_t}\sum_{i=1}^N \frac{2(N-i) + 1}{2N-1}R_i.
\label{eqn:hatB}
\end{equation}
We note that subscript $t$ refers to the tool surface, whereas $w$ is the deflection of the outermost layer (see Figure~\ref{fig:model}).
\subsubsection{Work done into foundation $U_F$}\label{sec:foundation}

\smallskip\noindent
The initial consolidation of the laminate $\eta$ from the debulk pressure, $q$, follows from inverting~\eqref{eqn:consolidation} such that $\eta = \sqrt{q/C}$. Differentiating \eqref{eqn:consolidation} with respect to $\eta$ gives the stiffness of the laminate at a given compressive strain, $\mathcal S(\eta) = 2C\eta$. The stiffness of the laminate in response to a wrinkle displacement $w$ about $\eta$, is found by expanding $\mathcal S$ at consolidation level $\eta - {w}/{Nt}$, yielding
\begin{equation}
\mathcal S\left(\eta - \frac{w}{Nt}\right) = 2C\left(\eta - \frac{w}{Nt}\right) = 2C\eta - 2\frac{C}{Nt}w.
\label{eqn:foundation}
\end{equation}
 We assume that a wrinkle deformation is resisted by a Winkler foundation which acts locally and normal to the ply, and therefore the force in response to a wrinkle displacement $w$ is given by
\begin{equation}
 f(w) = \mathcal S(\eta,w)w = 2C\eta w- \frac{2C}{Nt}w^2.
\label{eqn:forcefoundation}
\end{equation}
The total elastic strain energy stored in the foundation is
\begin{align}
U_F &=  \int_{-\frac{1}{2}\ell_1}^{\frac{1}{2}\ell_1} \frac{1}{2} 2C\eta w^2 - \frac{2}{3}\frac{C}{Nt} w^3\;dx_1 \\ \nonumber
&= \frac{R_1}{R_t}\int_{-\frac{1}{2}\ell_t}^{\frac{1}{2}\ell_t} 2C\eta w^2 - \frac{2}{3}\frac{C}{Nt} w^3\;dx_t.
\end{align}

\subsubsection{Inextensibility constraint}

\smallskip\noindent
The inextensibility constraint imposed on each layer is introduced as a constraint equation with an unknown Lagrange multiplier. Physically this can be seen as a rigid loading constraint, where the Lagrange multiplier is the load $p$, Fig.~\ref{fig:setup}. The length of outside layer is given by
\begin{equation}
\frac{1}{2}\pi R_i(0) = \int_{-\frac{1}{2}\ell_t}^{\frac{1}{2}\ell_t}\sqrt{1 + w^{\prime\; 2}}\;dx_1 = \frac{R_1}{R_t}\int_{-\frac{1}{2}\ell_t}^{\frac{1}{2}\ell_t}\sqrt{1 + w^{\prime\; 2}}\;dx_1.
\end{equation}
Applying equation~\eqref{eqn:lambda} and making the small displacement approximation $\sqrt{1 + w^{\prime\; 2}} \simeq 1 + \frac{1}{2}w^{\prime\; 2}$, the inextensibility constraint can be written as
\begin{equation}
\lambda_1  = \frac 12 \frac {R_1}{R_t}\int_{-\frac{1}{2}\ell_t}^{\frac{1}{2}\ell_t} w^{\prime\; 2}\;dx_t.
\label{eqn:constraint}
\end{equation}
The Lagrange multiplier $p$ which imposes this constraint is to be found as part of the solution.

\subsubsection{Total potential energy and Euler-Lagrange equation}

The total potential energy of the system is the sum of work done in bending and into the foundation, minus the Lagrange multiplier $p$ times the inextensibility constraint~\eqref{eqn:constraint}, such that
\begin{equation}
V =  \int^{\frac{1}{2}\ell_t}_{-\frac{1}{2}\ell_t} \frac12 \hat B w^{\prime\prime\; 2} -\frac 12 p\frac {R_1}{R_t}w^{\prime\; 2}+ \frac {R_1}{R_t}C\eta w^2 - \frac{2}{3}\frac {R_1}{R_t}\frac{C}{Nt} w^3\;dx_t.
\label{eqn:Energy}
\end{equation}

\smallskip\noindent
The total potential energy $V(w)$ is invariant under the reflective transformation, $(x,w)\mapsto (-x,w)$. This suggests the possibility of a symmetric mode as a minimising solution~\cite{DodwellSIAM}; therefore a symmetric section ($w^\prime = w^{\prime\prime\prime} = 0$) is imposed at $x_t = 0$. The correct boundary conditions at $x_t =\pm \ell_t$ are less clear, where neither clamped nor simply supported boundary conditions are physically relevant. The choice of homoclinic boundary conditions~\cite[Sec. 3.3]{Hunt2012b}, where $w$ and its derivatives vanish at $x_t =\pm \infty$, essentially removes the influence of the boundary on the solution, and suggests that a wrinkle forms over the corner radius and will decay to zero within the larger structure. Applying calculus of variations~\cite{Fox1963,Dodwellthesis2012} it follows that minimisers, $w$, of $V(w)$ are solutions of the Euler-Lagrange equation
\begin{equation}
\hat{B}w'''' + p\frac{R_1}{R_t}w'' + 2C\eta\frac{R_1}{R_t}w - \frac{2C}{Nt}\frac{R_1}{R_t} w^2 = 0,
\label{eq:EL}
\end{equation}
subject to symmetric and homoclinic boundary conditions, and the inextensibility constraint~\eqref{eqn:constraint}.
 
\section{Analytical and numerical solutions}

\subsection{Identification of modelling parameters}\label{sec:experiments}

All modelling parameters used for the results presented in the following section are summarised in Table \ref{tab:Properties};  these values are given for uncured M21-T700.

\begin{table*}
\begin{center}
\begin{tabular}{|c|c|c|c|c|c|c|}
\hline\centering
Parameter & $E_f$ (GPa) & $\phi_f$& $r_f$ ($\mu$m) & $t$(mm) & $R_t$ (mm) & $N$ \\ \hline
Value & 230 & 0.57 & 7.0 & 0.25 & 40 & 50 \\
Source & \cite{HexplyM21} & \cite{HexplyM21} & \cite{Wysocki2009} & \cite{HexplyM21} & - & - \\ \hline
\end{tabular}
\end{center}
\caption{Geometric and material parameters (per unit width) for M21-T700.}
\label{tab:Properties}
\end{table*}

\begin{figure*}
\centering
\includegraphics[width=2.7in]{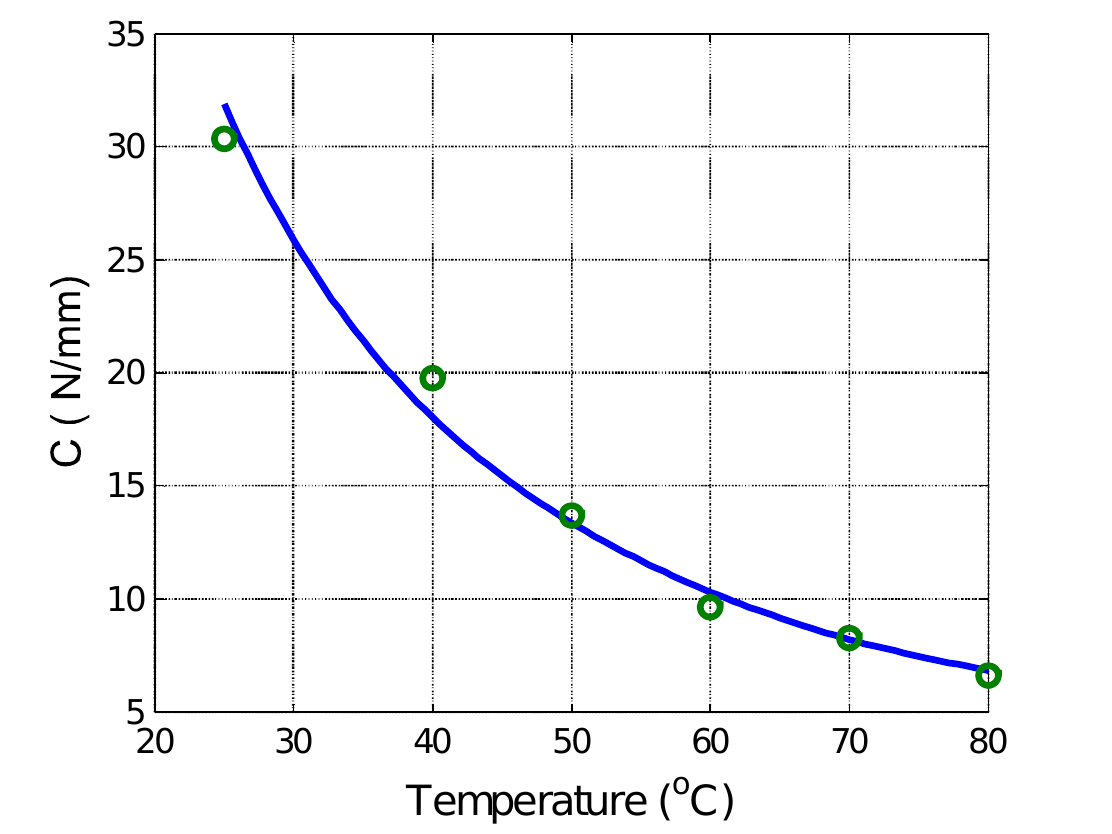} \includegraphics[width=2.7in]{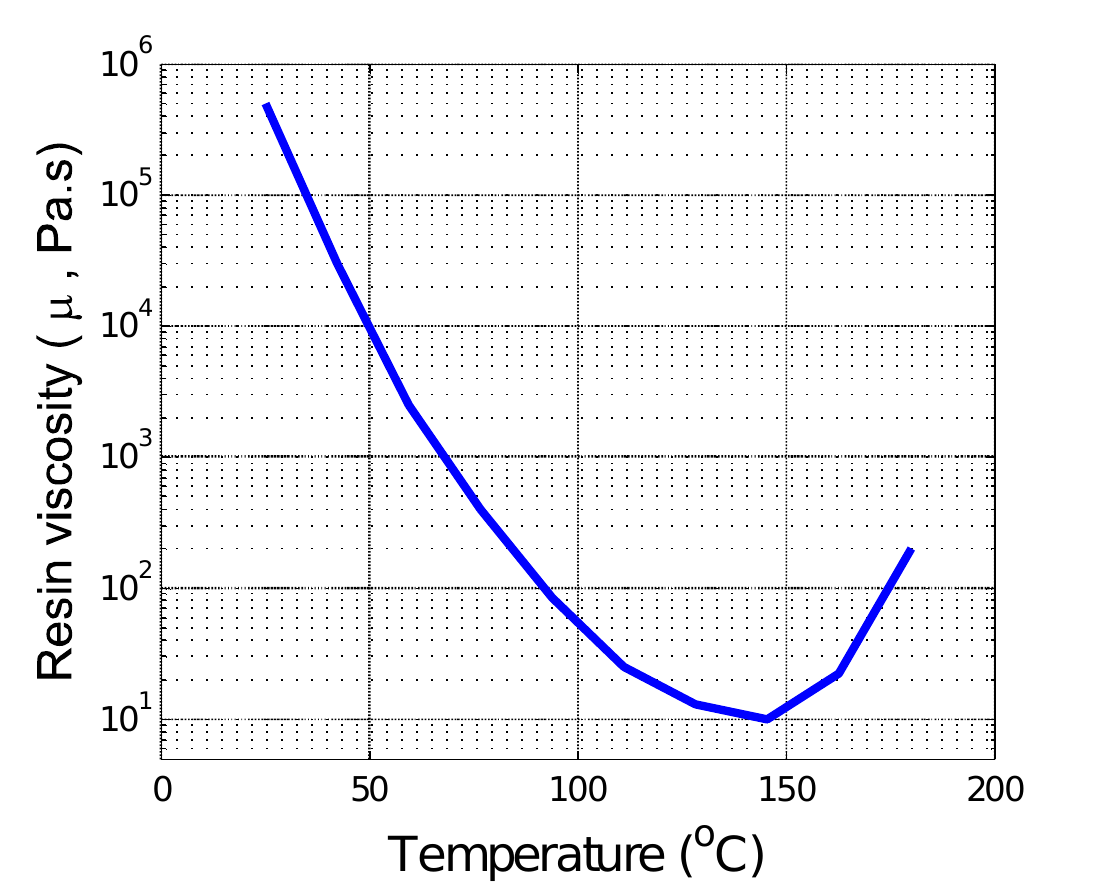}
\caption{ (Left) Plot of consolidation coefficient $C$ against temperature. (Right) Plot of resin viscosity against temperature, after~\cite{HexplyM21}.}
\centering
\label{fig:foundation}
\end{figure*}

The elastic behaviour of the lamina under compaction was characterized by consolidating six separate specimens at four distinct strains, with the experiments being repeated at six different temperatures. At each strain level, the displacement was held constant while the load settled to equilibrium. Using these relaxed loads the complete elastic response of the lamina is extracted. For this study, six cross-ply $[0,90]_{10}$ specimens of $50$mm$^2$ plies were compressed by an Instron-3369 within a heated oven, which controlled the compressive strain $\eta$ and temperature. The recorded loads were fitted to~\eqref{eqn:consolidation}, in the least-squared sense. Figure~\ref{fig:foundation} (left) shows a plot of consolidation coefficient $C$ against temperature. Finally Fig.~\ref{fig:foundation} (right) shows a plot of resin viscosity against temperature, provided by Hexcel~\cite{HexplyM21}.
\subsection{Critical wrinkling condition: A Galerkin approximation}\label{sec:criticalload}
The critical condition for a wrinkle corresponds to the linear buckling load of the system. Here we apply an energy based method~\cite{Thompson1973}, to find stationary solutions of the {\em linearised} total potential energy of the form $w(x) = A\cos\left(\pi x_t/\xi \right)$. Assuming this mode the {\em linearised} total potential energy is given by
\begin{equation}
V(\xi,A) = \frac{1}{2}\hat{B}\ell_t\left(\frac{\pi}{\xi}\right)^4A^2 - \frac{1}{2}\frac{R_1}{R_t}p\ell_t\left(\frac{\pi}{\xi}\right)^2A^2 + \frac{R_1}{R_t}C\eta\ell_t A^2.
\end{equation}
Stationary solutions of $V$ with respect to the amplitude $A$ are solutions of
\begin{equation}
\frac{\partial V}{\partial A} = A\ell_t\left(\hat{B}\left(\frac{\pi}{\xi}\right)^4 - p\frac{R_1}{R_t}\left(\frac{\pi}{\xi}\right)^2 + 2C\eta \frac{R_1}{R_t}\right) = 0
\end{equation}
Therefore either $A = 0$, the trivial flat state, or
\begin{equation}
p^c = \frac{R_t}{R_1}\frac{\hat{B}\pi^2}{\xi^2} + \frac{2C\eta\xi^2}{\pi^2}
\label{eqn:criticalcondition}
\end{equation}
and the system buckles. Finding the stationary value of this critical load with respect to the wavelength $\xi$, i.e solving $\partial p^c/\partial \xi = 0$ for $\xi$ it follows
\begin{equation}
\xi^{c} = \pi\sqrt[4]{\frac{\hat{B}{R_t}}{R_1 2C\eta}}.
\label{eqn:criticalwavelength}
\end{equation}

\subsection{Localisation due to the postbuckling asymmetry}\label{sec:localisation}

\begin{figure}
\centering
\includegraphics[width = 3in]{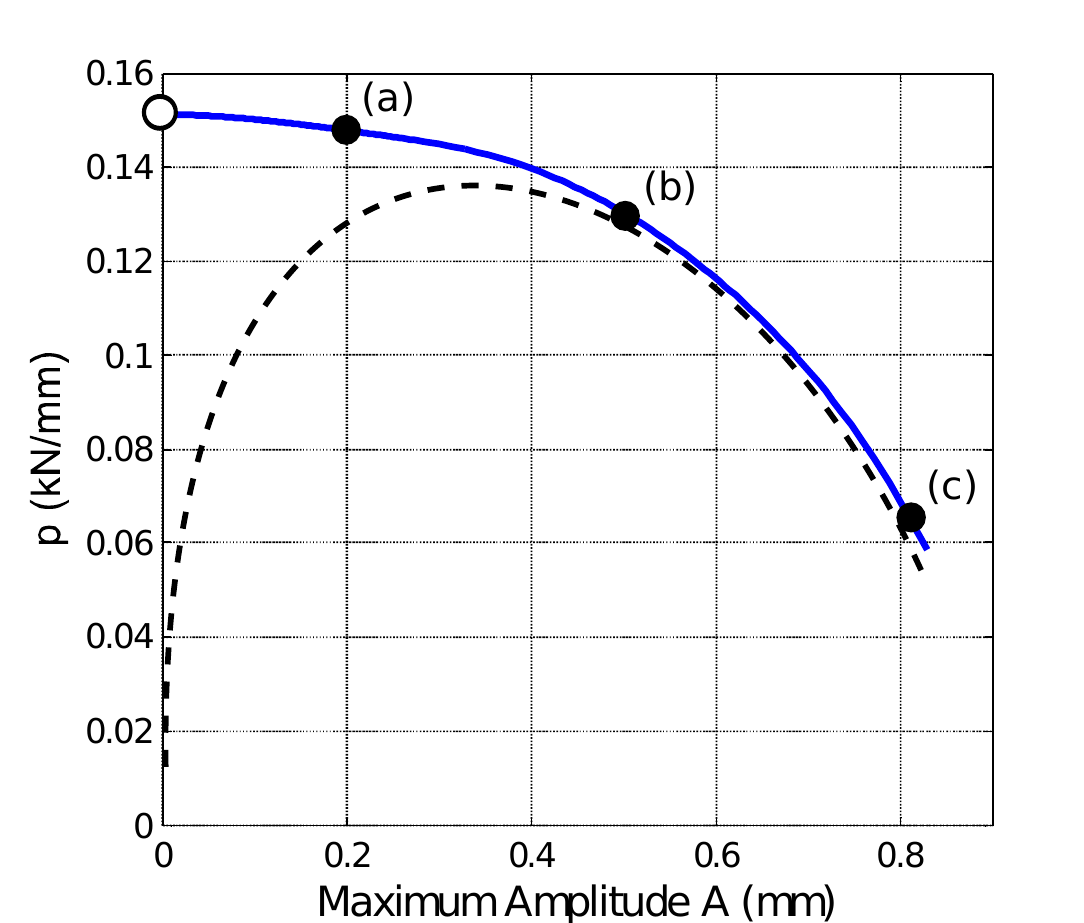}
\caption{Plot of maximum amplitude $A$ against load $p$ for numerical solutions of~\eqref{eq:EL} with modelling parameters taken at $25^{o}$C. Solid line shows {\em perfect} path where wrinkling bifurcation is marked with a hollow circle at $p^c = 0.15$kN/mm. The dashed lined shows an {\em imperfect} path found by introducing a small point load outwards at $x_t=0$. Solids circles indicate positions of solutions shown in Fig.~\ref{fig:solutions}.}
\label{fig:asymlocal}
\end{figure}

\begin{figure}
\centering
\includegraphics[width=3in]{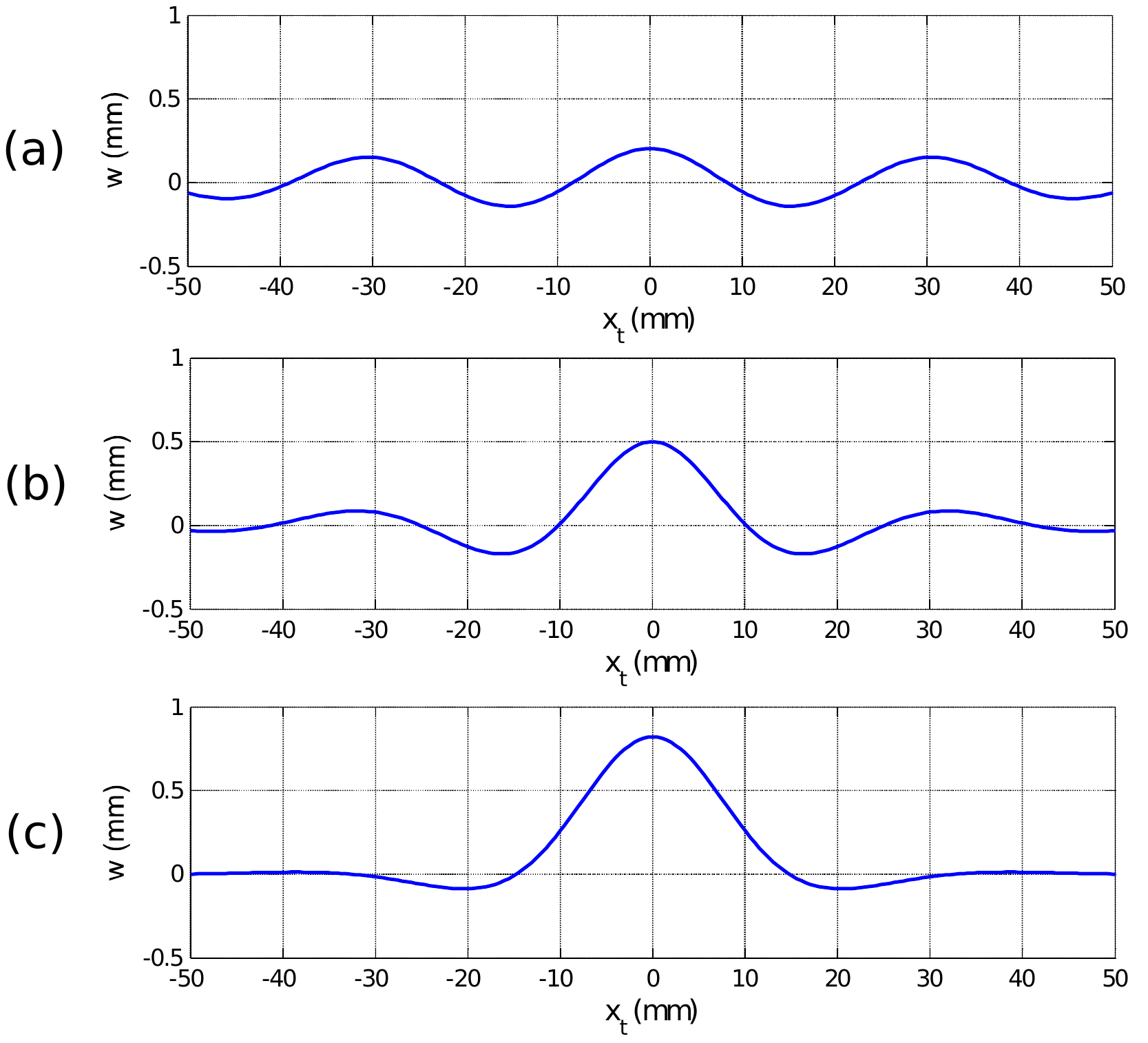}
\caption{Solutions, $w(x_t)$, of \eqref{eq:EL}, at (a) $A=0.20$mm (b) $A=0.50$mm and (c) $A=0.82$mm; for modelling parameters at $25^oC$.}
\label{fig:solutions}
\end{figure}

\smallskip\noindent
The solutions of the linearised problem correspond to periodic solutions over the complete domain. These are not observed in practice (Fig.~\ref{fig:introduction}), where micrographs of wrinkles demonstrate localised profiles.  We now observe how the nonlinear asymmetry gives rise to the possibility of localised fold solutions. The equation \eqref{eq:EL} is solved subject to constrained $\lambda$, where $p$ is to be found as part of the solution. Nonlinear solutions are tracked from the linear solution, $w(x) = A\cos(\pi x_t/\xi)$, into the post-buckled range using pseudo-arclength continuation~\cite{Seydel1988}. By rewriting~\eqref{eq:EL} and the length constraint~\eqref{eqn:constraint} as a system of five first-order equations~\cite{DodwellAccommodation2012}, the resulting boundary value problem is solved using the collocation method (e.g. \textsc{Matlab}'s solver \texttt{bvp4c}~\cite{Kierzenka2008}). Figure \ref{fig:asymlocal} shows the postbuckled path of the solutions, plotting the maximum amplitude $A(0)
$ against $p$. 
Numerical solutions at various 
postbuckled positions are given in Fig.~\ref{fig:solutions}.
\smallskip
The load-displacement plot shows an unstable postbuckled path, from the classical buckling instability at $p^c$. The system has two unstable subcritical postbuckling paths stemming from the bifurcation point, corresponding to positive or negative amplitude of the symmetric linear mode $w(x) = A\cos\left(\pi x_t/\xi \right)$. Figure~\ref{fig:asymlocal} shows the postbuckled branch from the {\em positive branch} which is of more physical relevance since it is of lower potential energy~\cite{HTB}. The imperfect paths, shown in Fig.~\ref{fig:asymlocal}, found by introducing an outward point load imperfection at $x_t = 0$, highlight that the system is particularly sensitive to imperfections. Therefore wrinkling/buckling instabilities may be triggered below the classical buckling load $p^c$.

\subsection{Critical limb lengths and wrinkle wavelengths}

\begin{figure*}
\centering
\includegraphics[width = 2.6in]{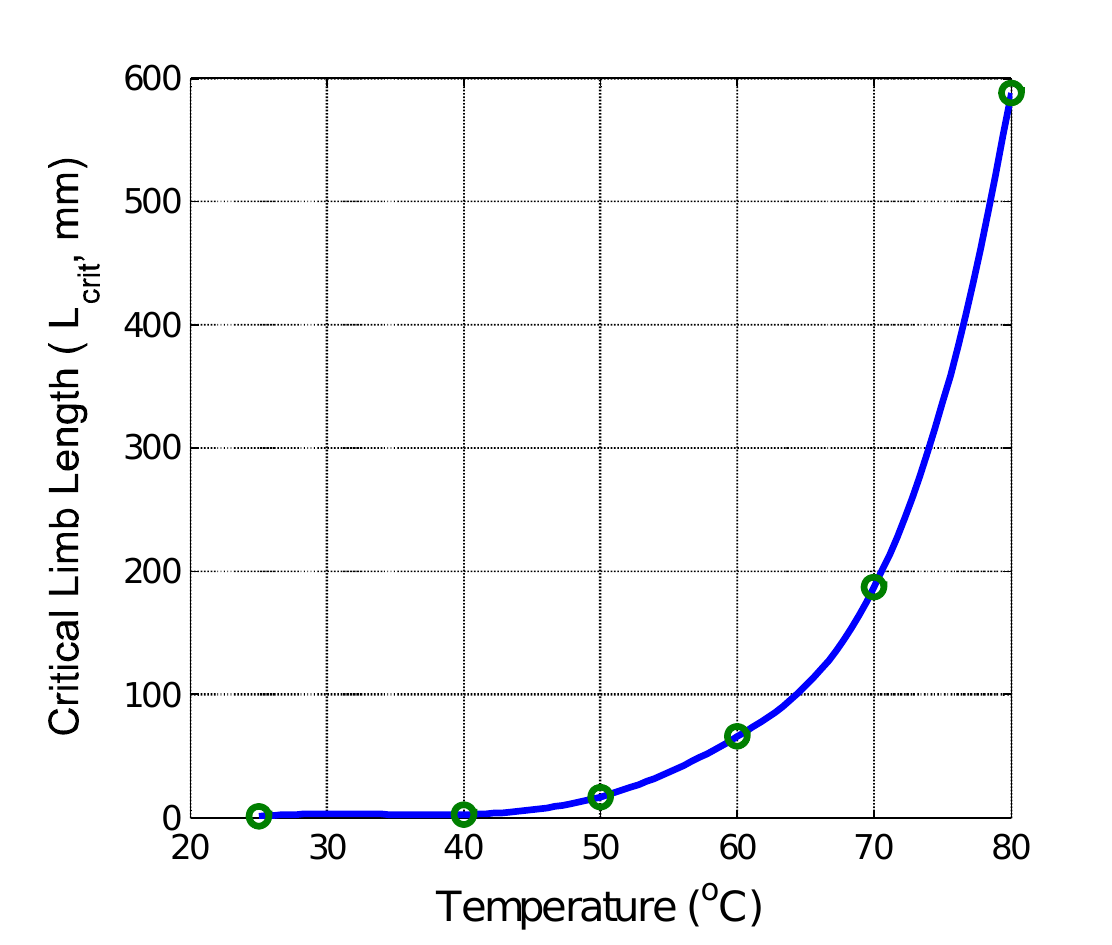}\includegraphics[width = 2.6in]{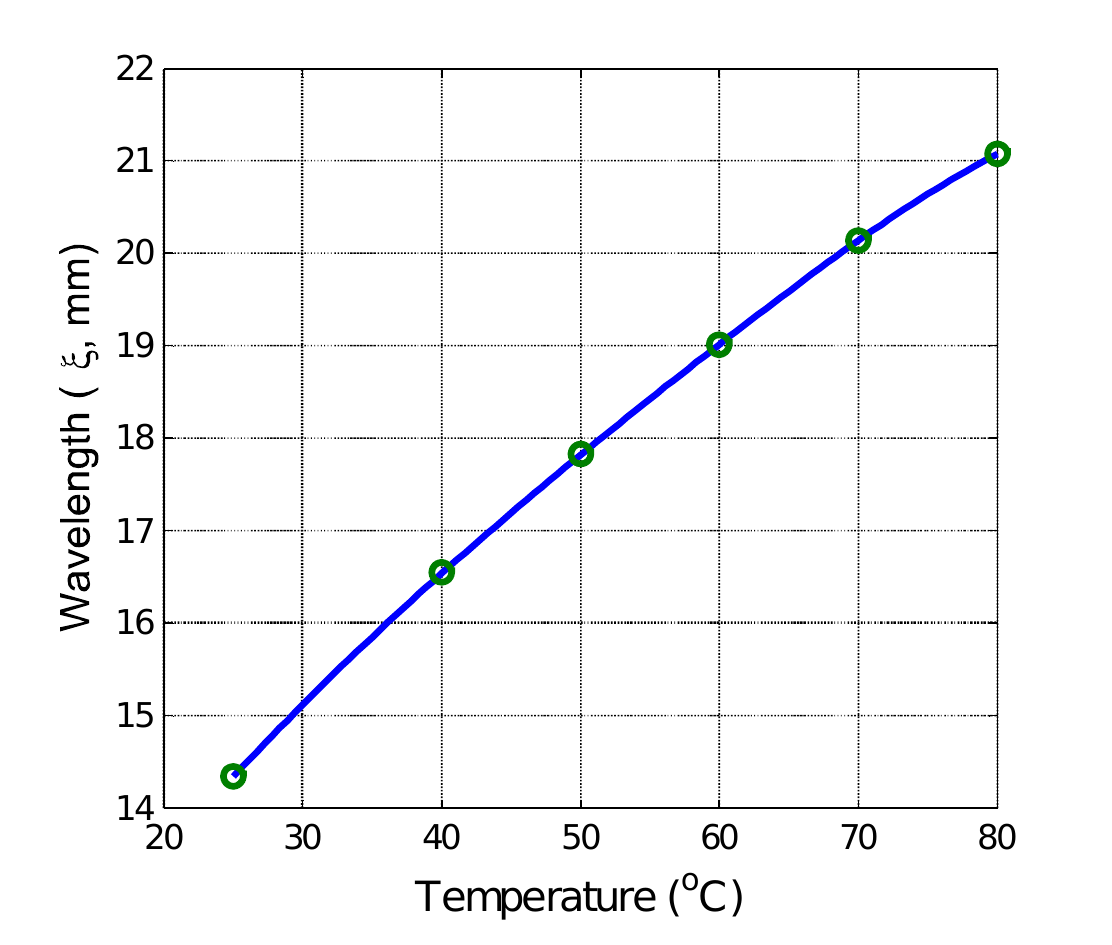}
\caption{For both cases a typical debulk time of $15$mins is assumed. (Left) Plot of critical limb length against temperature. (Right) Plot of wavelength $\xi$ against temperature. }
\label{fig:interplytemperature}
\end{figure*}

The model predicts that if the limb length $L> L_{crit}$ then a wrinkle will form. Since the postbuckling response is subcritical, $p < p_c$ for all non-trivial paths. As an approximation we have calculated the critical limb length $L_{crit}$ corresponding to the buckling load $p^c$. Figure~\ref{fig:interplytemperature} shows $L_{crit}$ against temperature for a typical debulk time of $T= 15$ mins, where other critical lengths for different debulk times ($T$) can be calculated by noting that $L_{crit}$ scales linearly with $T$~\eqref{eqn:yield}. The model predicts wrinkle wavelengths $\xi$ ranging from $14.33 - 21.07$mm, Figure~\ref{fig:interplytemperature} (right). The range of wavelengths show moderate agreement with the wrinkles observed in Fig.~\ref{fig:introduction} and similar micrographs, for which $\xi \simeq 13.5$mm. The predicted wavelengths are longer than those observed, which is due to assumptions imposed on the model. The restriction of the deformation modes through the laminate~\eqref{eqn:
lineardecay} limits the degrees of freedom by which the system can deform. As a result, the response is overly stiff, in particular leading to an over-estimate of the effective bending stiffness $\hat B$ and hence the wavelength~\eqref{eqn:criticalwavelength}.

\section{Discussion: strategies to minimise wrinkling}\label{sec:manufacturing}

For ambient debulks (e.g. $25^{o}$C) of $15$ minutes the predicted critical limb lengths are minimal, $L_{crit.} = 0.9$mm. The results show that for debulk temperatures up to $60^{o}$C laminates are particularly susceptible to wrinkling. On the other hand, wrinkles are unlikely to form at temperatures greater than $60^{o}C$, particular as the resin viscosity drops to $10 Pas$ at $145^oC$ during the cure cycle (Fig.~\ref{fig:foundation}) (right). However, intermediate ambient and heated debulks at lower temperatures may introduce significant imperfections, causing a wrinkle to manifest itself in the final cured part, Fig.~\ref{fig:introduction}. Various strategies can be adopted to encourage the formation of book-ends (and therefore avoiding wrinkling):
\begin{itemize}
 \item[\textbf{(A)}]\textbf{Choose material with low resin viscosity}. Current developments in carbon fibre composite technology mean that the new `no-flow' materials may be more susceptible to wrinkle defects. By restricting resin flow a consistent distribution of resin and fibre is maintained throughout a component, leading to significant strength and toughness benefits. However, these \textit{flow restrictors} significantly increase the resin viscosity, thereby limiting inhibiting the formation of {\em book-ends}.
 \item[\textbf{(B)}]\textbf{Heated debulks}. Figure~\ref{fig:interplytemperature} may be used to assess the minimum temperature
required for debulks..
\end{itemize}
Aside from material selection other strategies can be adopted to minimize the probability of wrinkles forming. Simply reducing the required length change of the each ply will decrease the require shear strain over the limbs. This can be achieved in several ways:
\begin{itemize}
 \item[\textbf{(C)}] \textbf{Reduce the severity of the geometry}. For this corner radius this is equivalent to increasing the tool radius $R_t$. For a tapered section this might be the ramp rate or for more complex three-dimensional features the Gaussian curvature. In the limit of $R_t \rightarrow \infty$ the surface is flat and no length change is required. The choice of tool geometry is a compromise between design and ease of manufacture.
 
 \item[\textbf{(D)}] \textbf{Multiple intermediate debulks}. The amount the plies have to slip over one another during a single debulk, can be reduced by including multiple debulks in the laying up process. This effectively increases the tool radius and reduces the amount of consolidation at each debulk.
\end{itemize}
The second of these solutions is particularly costly, by adding significant lead time to the manufacture of large components. Therefore a wrinkle-free process, requiring minimal intermediate debulks is a significant commercial challenge. An interesting avenue of future research will consider the possibility of more {\em  exotic} layup sequences to aid manufacturiability.
\begin{itemize}
 \item[\textbf{(E)}] \textbf{Layup optimisation for manufacturability}. In the work presented it is assumed that that fibres are inextensible and align with the principal compressive stress generated during consolidation, making the laminate  particularly susceptible buckling/wrinkling stabilities. Layup strategies could be adopted to reduce this effect by choosing ply orientations for the outermost plies which do not align with this compressive stress. If the layup sequence was constrained to be symmetric, then fibres which are oriented in this principal direction would be best placed close to the middle of the laminate. However unsymmetric layups, may offer a direct strategy to reduce wrinkles. 
\end{itemize}

\section{Concluding remarks}

This paper presents a simplified one-dimensional model which encapsulates the elastic buckling/wrinkling response of fibres as they simultaneously consolidated over a radius whilst being constrained axially. The work highlights the importance of including the individual contribution of a layer to bending. If classical laminate continuum models are adopted, the stiffness of the laminate level response swamps the localised behaviour of individual plies and incorrect results are achieved. Once a ply has the freedom to bend and slip independently, they become particularly susceptible to localised buckling instabilities. This tendency to localise is emphasised by the combined effects of long thin layers and the asymmetric nature of the consolidation process. Whilst there is scope to extend this model to account for more complex nonlinear viscoelastic effects, from a manufacturing perspective the analysis provides indications of possible strategies  which may reduce the chance of wrinkles forming. Perhaps most 
importantly, the model shows that if layers cannot slip wrinkles are very likely to form. Therefore manufacturing processes and materials should be selected to allow layers to accommodate complex geometries. 

A vital extension to this work will be to consider the consolidation over three-dimensional geometries. The directional nature of different plies throughout a laminate means that in directions orthogonal to the fibre directions these plies may offer little or no bending stiffness, yet may be heavily constrained by consolidation in the perpendicular direction. These interactions will require the careful development of new analytical tools. We believe these are key factors in the fully understanding the formation wrinkling of carbon fibre composites during manufacture of large components.

\section*{Acknowledgements}
The authors would like to acknowledge GKN Aerospace and TSB for supporting this work under the grant TSB:1000774. We are grateful to Ian Lang, Christopher Jones and Richard Newley, from GKN Aerospace, for many useful discussions.

\bibliographystyle{elsart-num}
\bibliography{library}

\end{document}